\documentclass[conference]{IEEEtran}
\IEEEoverridecommandlockouts
\usepackage{cite}
\usepackage{amsmath,amssymb,amsfonts}
\usepackage{algorithmic}
\usepackage{algorithm}
\usepackage{graphicx}
\usepackage{textcomp}
\usepackage{xcolor}
\usepackage{stfloats}
\def\BibTeX{{\rm B\kern-.05em{\sc i\kern-.025em b}\kern-.08em
    T\kern-.1667em\lower.7ex\hbox{E}\kern-.125emX}}

\usepackage{geometry}
\renewcommand{\baselinestretch}{0.98}

\begin{document}

\title{\huge Movable-Antenna Array Enhanced Downlink NOMA }

\author{
	\IEEEauthorblockN{Nianzu Li\IEEEauthorrefmark{2},~Peiran Wu\IEEEauthorrefmark{2},~Lipeng Zhu\IEEEauthorrefmark{4},~and~Derrick Wing Kwan Ng\IEEEauthorrefmark{3}}
	\IEEEauthorblockA{\IEEEauthorrefmark{2}School of Electronics and Information Technology, Sun Yat-sen University, Guangzhou, China 510006}
	\IEEEauthorblockA{\IEEEauthorrefmark{4}Department of Electrical and Computer Engineering, National University of Singapore, Singapore 117583}
	\IEEEauthorblockA{\IEEEauthorrefmark{3}School of Electrical Engineering and Telecommunications, University of New South Wales, Sydney, NSW2052, Australia}
	\IEEEauthorblockA{E-mail: linz5@mail2.sysu.edu.cn, wupr3@mail.sysu.edu.cn, zhulp@nus.edu.sg, w.k.ng@unsw.edu.au}
}

\maketitle

\begin{abstract}
Movable antenna (MA) has gained increasing attention in the field of wireless communications due to its exceptional capability to proactively reconfigure wireless channels via localized antenna movements. In this paper, we investigate the resource allocation design for an MA array-enabled base station serving multiple single-antenna users in a downlink non-orthogonal multiple access (NOMA) system. We aim to maximize the sum rate of all users by jointly optimizing the transmit beamforming and the positions of all MAs at the BS, subject to the constraints of transmit power budget, finite antenna moving region, and the conditions for successive interference cancellation decoding rate. The formulated problem, inherently highly non-convex, is addressed by successive convex approximation (SCA) and alternating optimization methods to obtain a high-quality suboptimal solution. Simulation results unveil that the proposed MA-enhanced downlink NOMA system can significantly improve the sum rate performance compared to both the fixed-position antenna (FPA) system and the traditional orthogonal multiple access (OMA) system.
\end{abstract}


\section{Introduction}
With rapid advancements in wireless communication technologies, global anticipation for the upcoming sixth-generation (6G) networks has dramatically increased. Specifically, these networks are poised to support high data rates, reduced latency, massive access, and simultaneous connectivity for a vast number of devices. However, conventional communication systems with fixed-position antennas (FPAs) face severe limitations in fulfilling these requirements due to their fixed and discrete antenna layouts, lacking the ability to fully exploit continuous channel variations in spatial regions\cite{ref1}. Fortunately, these limitations are expected to be addressed with the emergence of movable antenna (MA) technology\cite{ref3,ref35}. Unlike conventional FPA systems, MA-enabled systems can adaptively reposition the transmit/receive antennas to reconfigure  wireless channels to cope with the dynamic nature of radio environments, thus providing additional degrees of freedom (DoFs) to improve system performance.

Recently, the great benefits of MAs have driven numerous research efforts into integrating them into various wireless systems. For instance, the authors in \cite{ref27} investigated an MA-assisted multiple-input single-output (MISO) single-user communication system and proposed a graph-based approach to obtain the optimal MA positions to maximize the received signal's power. Besides, in \cite{ref9}, the authors investigated multi-user communications enhanced by MAs, where the power consumption of uplink transmissions from multiple single-MA users to an FPA-enabled base station (BS) was minimized by jointly designing the positions of MAs, the transmit powers of users, and the receive combining matrix of the BS. Furthermore, the authors in \cite{ref39} investigated multicast communication systems aided by both transmit and receive MAs, where the minimum weighted signal-to-interference-plus-noise ratio (SINR) among all users was maximized by jointly optimizing all MAs' positions and the beamforming vectors. In addition, MAs have also been integrated to other communication scenarios, such as over-the-air computing\cite{ref36}, covert communications\cite{ref43}, and intelligent reflecting surface-aided communications\cite{ref37}.

On the other hand, non-orthogonal multiple access (NOMA) also emerges as one of the potential technologies for future wireless communication networks, which enables distinct users to share the same time/frequency/code resource block for enhancing the spectral efficiency and system throughput\cite{ref38}. Therefore, integrating MA into NOMA systems holds significant potential to further improve system performance by synergizing their advantages. To this end, some existing works have demonstrated the superior performance of combining MA with NOMA over conventional FPA-based NOMA and orthogonal multiple access (OMA), e.g., \cite{ref41,ref2,ref42}. However, only a single MA is considered to be deployed at the user side in these works, while there are typically more space and energy available at the BS side for antenna movement. To fill in this gap, in this paper, we explore a more general scenario for the MA-enhanced donwlink NOMA system, where multiple MAs are deployed at the BS side to serve multiple single-FPA users. Specifically, we formulate an optimization problem for maximizing the users' sum rate by jointly optimizing the beamforming vectors of the BS and the antenna positioning vector (APV) of all MAs. To tackle the non-convex nature of this problem, we propose an iterative algorithm capitalizing on alternating optimization and successive convex approximation (SCA) techniques to acquire a suboptimal solution. Simulation results confirm that the proposed MA-enhanced downlink NOMA system outperforms both the FPA system and the traditional OMA system. 



\section{System Model}
\begin{figure}[t]
	\centering
	\includegraphics[width=0.469\textwidth]{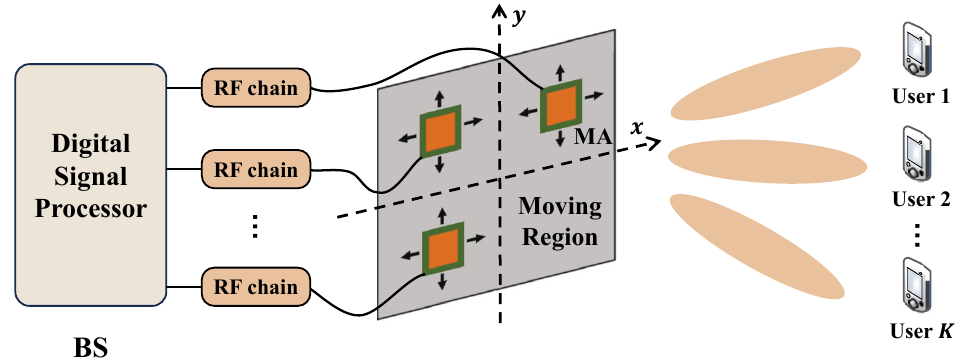}
	\caption{Illustration of the MA-enhanced downlink MISO-NOMA system.}
	\label{system_model}
	\vspace{-0.2cm}
\end{figure}
As shown in Fig. 1, we consider an MA-enhanced downlink MISO system employing NOMA, where a BS is equipped with $M$ transmit MAs to serve $K$ single-antenna users. The set of all users is denoted by $\mathcal{K}=\left\{1,2,\dots,K\right\}$ and the set of all MAs is denoted by $\mathcal{M}=\left\{1,2,\dots,M\right\}$. Each MA at the BS is connected to
the radio frequency (RF) chain via a flexible cable and thus can be repositioned in a local two-dimensional (2D) region, $\mathcal{C}_t$, for proactively improving channel conditions\cite{ref1,ref3}. The position of the $m$-th MA is represented by the 2D Cartesian coordinates, i.e., $\mathbf{u}_m=\left[x_m,y_m\right]^{\mathrm{T}}\in\mathcal{C}_t,\forall m\in\mathcal{M}$. Without loss of generality, we assume that $\mathcal{C}_t$ is a square region with side length $A$.

Let $\mathbf{h}_k(\tilde{\mathbf{u}})\in\mathbb{C}^{M\times1}$ denote the channel vector between the $M$ MAs at the BS and user $k$, with $\tilde{\mathbf{u}}=\left[\mathbf{u}_1^{\mathrm{T}},\mathbf{u}_2^{\mathrm{T}},\dots,\mathbf{u}_M^{\mathrm{T}}\right]^{\mathrm{T}}$ denoting the APV for all MAs. Therefore, the received signal at user $k$ can be expressed as
\begin{align}
	y_k=\mathbf{h}_k^{\mathrm{H}}(\tilde{\mathbf{u}})\sum_{j=1}^{K}\mathbf{w}_js_j+n_k,
\end{align}
where $s_j\in\mathbb{C}$ denotes the information-bearing symbol for user $j$ with zero mean, unit variance, and $\mathbb{E}\left\{s_j^*s_q\right\}=0,\forall j\neq q$. Also, $\mathbf{w}_j$ represents the corresponding beamforming vector and $n_k\sim\mathcal{CN}\left(0,\sigma_k^2\right)$ is the zero-mean additive white Gaussian
noise (AWGN) at user $k$ with power $\sigma_k^2$. 

\subsection{Channel Model}
In this paper, we employ the field-response based channel model under the far-field condition\cite{ref3}, where the channel response from the BS to each user is the superposition of the coefficients of multiple channel paths. Let $L_k,\forall k \in \mathcal{K}$, denote the number of transmit channel paths from the BS to user $k$. The elevation and azimuth angles of departure (AoDs) for the $\ell$-th transmit path between the BS and user $k$ are denoted as $\theta_{k,\ell}$ and $\phi_{k,\ell}$, respectively. Then, the transmit field-response vector (FRV) for the channel between the $m$-th MA at the BS and user $k$ can be given by
\begin{align}
	\mathbf{g}_k(\mathbf{u}_m)=\left[e^{\mathrm{j}\frac{2\pi}{\lambda}\rho_{k,1}(\mathbf{u}_m)},\dots,e^{\mathrm{j}\frac{2\pi}{\lambda}\rho_{k,L_k}(\mathbf{u}_m)}\right]^{\mathrm{T}},
\end{align}
where $\lambda$ denotes the signal carrier wavelength and $\rho_{k,\ell}(\mathbf{u}_m)=x_m\sin\theta_{k,\ell}\cos\phi_{k,\ell}+y_m\cos\theta_{k,\ell},1\leq\ell\leq L_k$, denotes the signal propagation difference for the $\ell$-th channel path of user $k$ between the $m$-th MA's position $\mathbf{u}_m$ and the reference point at the BS, i.e., $\mathbf{o}=[0,0]^{\mathrm{T}}$. As a result, the channel vector between the BS and user $k$ can be obtained as
\begin{align}
	\mathbf{h}_k(\tilde{\mathbf{u}})=\mathbf{G}_k^{\mathrm{H}}(\tilde{\mathbf{u}})\mathbf{f}_k,
\end{align}
where $\mathbf{G}_k(\tilde{\mathbf{u}})=\left[\mathbf{g}_k(\mathbf{u}_1),\dots,\mathbf{g}_k(\mathbf{u}_M)\right]\in\mathbb{C}^{L_k\times M}$ denotes the field-response matrix (FRM) for all MAs at the BS, and $\mathbf{f}_k=\left[f_{k,1},f_{k,2},\dots,f_{k,L_k}\right]^{\mathrm{T}}\in\mathbb{C}^{L_k\times 1}$ denotes the path-response vector (PRV), representing the $L_k$ multi-path coefficients from the reference point at the BS to user $k$.

\subsection{Problem Formulation}
According to the NOMA protocol\cite{ref38}, each user applies successive interference cancellation (SIC) to recover its signal. Let $s(k)$ denote the decoding order of user $k$. If $s(k)=i$, the signal of user $k$ will be the $i$-th one to be decoded. As a result, the signal-to-interference-plus-noise ratio (SINR) of user $k$ to decode its own signal is given by
\begin{align}
	\gamma_{k\rightarrow k}=\frac{\left|\mathbf{h}_k^{\mathrm{H}}\left(\tilde{\mathbf{u}}\right)\mathbf{w}_k\right|^2}{\sum_{s(j)>s(k)}\left|\mathbf{h}_k^{\mathrm{H}}\left(\tilde{\mathbf{u}}\right)\mathbf{w}_j\right|^2+\sigma_k^2}.
\end{align}
The corresponding achievable rate of user $k$ to decode its own signal is thus given by 
$R_{k\rightarrow k}=\log_2\left(1+\gamma_{k\rightarrow k}\right)$.
Besides, for any $s(k)<s(\bar{k})$, the SINR of user $\bar{k}$ to decode the signal for user $k$ is given by
\begin{align}
	\gamma_{k\rightarrow \bar{k}}=\frac{\left|\mathbf{h}_{\bar{k}}^{\mathrm{H}}\left(\tilde{\mathbf{u}}\right)\mathbf{w}_k\right|^2}{\sum_{s(j)>s(k)}\left|\mathbf{h}_{\bar{k}}^{\mathrm{H}}\left(\tilde{\mathbf{u}}\right)\mathbf{w}_j\right|^2+\sigma_{\bar{k}}^2},
\end{align}
based on which the achievable rate of user $\bar{k}$ to decode the signal for user $k$ can be given by 
$R_{k\rightarrow \bar{k}}=\log_2\left(1+\gamma_{k\rightarrow \bar{k}}\right)$. Note that to ensure that the SIC can be performed successfully, the achievable rate of user $\bar{k}$ to decode user $k$'s signal should be no less than that of user $k$ to decode its own signal, i.e., 
$R_{k\rightarrow k}\leq R_{k\rightarrow \bar{k}},~\mathrm{if~} s(k)< s(\bar{k})$.

In this paper, we aim to maximize the sum rate of all users by jointly optimizing the APV for all MAs, the beamforming vectors at the BS, and the SIC decoding order. To investigate the performance limit, we assume that all involved channel state information (CSI) is perfectly known at the BS side for resource allocation design. Then, the optimization problem can be formulated as
\begin{subequations}
	\begin{align}
		\max_{\tilde{\mathbf{u}},\{\mathbf{w}_k,s(k)\}}& \quad \sum_{k=1}^{K}R_{k\rightarrow k}\\
		\mathrm{s.t.}
		& \quad \mathbf{u}_m\in\mathcal{C}_t,~\forall m \in \mathcal{M},\\
		& \quad \|\mathbf{u}_m-\mathbf{u}_n\|_2\geq D,~\forall m,n\in\mathcal{M},m\neq n,\\
		& \quad R_{k\rightarrow k}\geq R_k^{\min},~\forall k \in \mathcal{K},\\
		& \quad R_{k\rightarrow k}\leq R_{k\rightarrow \bar{k}},~s(k)< s(\bar{k}),\\
		& \quad \sum_{k=1}^{K}\|\mathbf{w}_k\|^2\leq P_{s},\\
		& \quad s(k)\in\Omega,~\forall k\in\mathcal{K},
	\end{align}
\end{subequations}
where $D$ in constraint (6c) is the minimum inter-MA distance to avoid coupling effects among antennas; $R_k^{\min}$ in constraint (6d) is the minimum rate requirement for user $k$ to guarantee its quality-of-service (QoS); $P_s$ in constraint (6f) is the total transmit power at the BS, and $\Omega$ in constraint (6g) is the set of all possible SIC decoding orders. Problem (6) is challenging to solve optimally since the objective function is highly non-concave with respect to $\tilde{\mathbf{u}},\{\mathbf{w}_k\}$, and $\{s(k)\}$. Besides, the three high-dimensional optimization variables are highly coupled with each other, rendering the problem more intractable.

\section{Proposed Solution}
\subsection{Problem Transformation}
Since the decoding orders $\{s(k)\}$ in problem (6) are combinatorial optimization variables, i.e., the number of all possible decoding orders is finite, we can determine the optimal sum rate by first solving problem (6) with all decoding orders and then selecting the one that yields the maximum sum rate. Then, by introducing two set of slack optimization variables, $\{\alpha_{k,i}\}$ and $\{\beta_{k,i}\}$, problem (6) with a given decoding order can be equivalently transformed into\cite[Proposition 1]{ref38}
\begin{subequations}
	\begin{align}
		\max_{\tilde{\mathbf{u}},\{\mathbf{w}_k,R_{k\rightarrow i},\alpha_{k,i},\beta_{k,i}\}}& \quad \sum_{k=1}^{K}R_{k\rightarrow k}\\
		\mathrm{s.t.}
		& \quad R_{k\rightarrow i}\leq \log_2\left(1+\frac{1}{\alpha_{k,i}\beta_{k,i}}\right),\notag\\
		&\quad ~\forall k,i\in\mathcal{K},~ s(k)\leq s(i),\\
		& \quad \frac{1}{\alpha_{k,i}}\leq \left|\mathbf{h}_i^{\mathrm{H}}\left(\tilde{\mathbf{u}}\right)\mathbf{w}_k\right|^2,~\forall k,i\in\mathcal{K},\\
		& \quad \beta_{k,i}\geq\sum_{s(j)>s(k)}\left|\mathbf{h}_i^{\mathrm{H}}\left(\tilde{\mathbf{u}}\right)\mathbf{w}_j\right|^2+\sigma_i^2,\notag\\
		& \quad ~\forall k,i\in\mathcal{K},\\
		& \quad R_{k\rightarrow i}\geq\log_2\left(1+\frac{1}{\alpha_{k,k}\beta_{k,k}}\right),\notag\\
		&\quad ~\forall k,i\in\mathcal{K},~ s(k)< s(i),\\
		& \quad \mathrm{(6b),(6c),(6d),(6f)}.\notag
	\end{align}
\end{subequations}
However, problem (7) is still highly non-convex due to the intricate coupling variables. To resolve this issue, in the following subsection, we will introduce an alternating optimization framework to decompose problem (7) into several subproblems and address them efficiently by exploiting the SCA technique. 

\subsection{Alternating Optimization}
\textit{1) Optimizing $\{\mathbf{w}_k\}$ with given $\tilde{\mathbf{u}}$:} First, let us define $\mathbf{W}_k=\mathbf{w}_k\mathbf{w}_k^{\mathrm{H}},\forall k \in \mathcal{K}$. With any given APV $\tilde{\mathbf{u}}$, problem (7) can be transformed into
\begin{subequations}
	\begin{align}
		\max_{\{\mathbf{W}_k,R_{k\rightarrow i},\alpha_{k,i},\beta_{k,i}\}}& \quad \sum_{k=1}^{K}R_{k\rightarrow k}\\
		\mathrm{s.t.}
		& \quad \frac{1}{\alpha_{k,i}}\leq\mathrm{Tr}\left(\mathbf{W}_k\mathbf{h}_i\mathbf{h}_i^{\mathrm{H}}\right),~\forall k,i\in\mathcal{K},\\
		& \quad \beta_{k,i}\geq\sum_{s(j)>s(k)}\mathrm{Tr}\left(\mathbf{W}_j\mathbf{h}_i\mathbf{h}_i^{\mathrm{H}}\right)+\sigma_i^2,\notag\\
		& \quad ~\forall k,i\in\mathcal{K},\\
		& \quad \sum_{k=1}^{K}\mathrm{Tr}\left(\mathbf{W}_k\right)\leq P_s,\\
		& \quad \mathbf{W}_k\succeq\mathbf{0},~\forall k \in\mathcal{K},\\
		& \quad \mathrm{rank}(\mathbf{W}_k)=1,~\forall k \in\mathcal{K},\\
		& \quad \mathrm{(7b),(7e),(6d)}.\notag
	\end{align}
\end{subequations}
Note that problem (8) is still non-convex due to non-convex constraints (8f) and (7b). According to Theorem 1 in \cite{ref38}, the optimal solution to problem (8) without constraint (8f), i.e., semi-definite relaxation, can always satisfy $\mathrm{rank}(\mathbf{W}_k)=1,\forall k\in\mathcal{K}$. Therefore, we can drop the rank-one constraint (8f) without losing its optimality. Furthermore, since the right-hand-side (RHS) in constraint (7b) is jointly convex with respect to $\alpha_{k,i}$ and $\beta_{k,i}$, by applying the first-order
Taylor series expansion, we can derive the following lower bound on $\log_2\left(1+\frac{1}{\alpha_{k,i}\beta_{k,i}}\right)$ at any given local points $\left\{\alpha_{k,i}^{t},\beta_{k,i}^{t}\right\}$:
\begin{align}
	\log_2\left(1+\frac{1}{\alpha_{k,i}\beta_{k,i}}\right)&\geq\log_2\left(1+\frac{1}{\alpha_{k,i}^{t}\beta_{k,i}^{t}}\right)\notag\\
	&-\frac{\log_2e}{\alpha_{k,i}^t+{\alpha_{k,i}^t}^2\beta_{k,i}^t}\left(\alpha_{k,i}-\alpha_{k,i}^t\right)\notag\\
	&-\frac{\log_2e}{\beta_{k,i}^t+{\beta_{k,i}^t}^2\alpha_{k,i}^t}\left(\beta_{k,i}-\beta_{k,i}^t\right)\notag\\
	&\triangleq \vartheta_{k,i}^t\left(\alpha_{k,i},\beta_{k,i}\right).
\end{align}
As a result, in the $t$-th iteration of SCA, a convex subset of constraint (7b) can be established as
\begin{align}
	R_{k\rightarrow i}\leq\vartheta_{k,i}^t\left(\alpha_{k,i},\beta_{k,i}\right),~\forall k,i\in\mathcal{K},~ s(k)\leq s(i),
\end{align}
based on which the optimization problem is transformed into
\begin{subequations}
	\begin{align}
		\max_{\{\mathbf{W}_k,R_{k\rightarrow i},\alpha_{k,i},\beta_{k,i}\}}& \quad \sum_{k=1}^{K}R_{k\rightarrow k}\\
		\mathrm{s.t.}
		& \quad \mathrm{(8b)-(8e),(6d),(7e),(10)}.\notag
	\end{align}
\end{subequations}
Problem (11) is a semidefinite programming (SDP) problem and thus can be optimally solved via the CVX toolbox\cite{ref32}. Denoting the optimal solution to problem (11) as $\{\mathbf{W}_k^{\star}\}$, the optimal beamforming vector at the BS, i.e., $\left\{\mathbf{w}_k^{\star}\right\}$, can be obtained from $\{\mathbf{W}_k^{\star}\}$ through eigenvalue decomposition.

\textit{2) Optimizing $\mathbf{u}_m$ with given $\left\{\mathbf{w}_k\right\}$ and $\left\{\mathbf{u}_n,n\neq m\right\}$:} With any given beamforming vectors $\left\{\mathbf{w}_k\right\}$ and antenna positions $\left\{\mathbf{u}_n,n\neq m\right\}$, problem (7) is simplified as
\begin{subequations}
	\begin{align}
		\max_{\mathbf{u}_m,\{R_{k\rightarrow i},\alpha_{k,i},\beta_{k,i}\}}& \quad \sum_{k=1}^{K}R_{k\rightarrow k}\\
		\mathrm{s.t.}
		& \quad \mathrm{(6b)-(6d),(7b)-(7e)}.\notag
	\end{align}
\end{subequations}
Problem (12) is a non-convex problem due to the non-convex constraints (6c), (7b), (7c), and (7d), rendering it challenging to acquire the optimal solution. Similarly, we utilize the SCA
technique to address the problem. 

In the previous discussion,
we have showed that (10) is a convex subset of non-convex constraint (7b). Next, for handling non-convex constraint (7c), let us first define $\Gamma_{k,i}(\mathbf{u}_m)\triangleq\left|\mathbf{h}_i^{\mathrm{H}}\left(\tilde{\mathbf{u}}\right)\mathbf{w}_k\right|^2,\forall k,i$. Then, denote the gradient and Hessian matrix of $\Gamma_{k,i}(\mathbf{u}_m)$ over $\mathbf{u}_m$ as $\nabla\Gamma_{k,i}(\mathbf{u}_m)$ and $\nabla^2\Gamma_{k,i}(\mathbf{u}_m)$, respectively, with their derivations provided in Appendix A.
Drawing on Taylor’s theorem,
we can construct a quadratic concave function to globally lower-bound $\Gamma_{k,i}(\mathbf{u}_m)$, i.e., the RHS of (7c), denoted as
\begin{align}
	\Gamma_{k,i}\left(\mathbf{u}_m\right)\geq&\Gamma_{k,i}\Big(\mathbf{u}_m^t\Big)+\nabla\Gamma_{k,i}\Big(\mathbf{u}_m^t\Big)^\mathrm{T}\Big(\mathbf{u}_m-\mathbf{u}_m^t\Big)\notag\\
	&-\frac{\delta_{k,i}}{2}\Big(\mathbf{u}_m-\mathbf{u}_m^t\Big)^{\mathrm{T}}\Big(\mathbf{u}_m-\mathbf{u}_m^t\Big)\notag\\
	\triangleq&\Gamma_{k,i}^{\mathrm{lb},t}\left(\mathbf{u}_m\right),
\end{align}
where $\delta_{k,i}$ is a positive real number satisfying $\delta_{k,i}\mathbf{I}_2\succeq\nabla^2\Gamma_{k,i}(\mathbf{u}_m)$. 
Note that since we have $\left\|\nabla^2\Gamma_{k,i}(\mathbf{u}_m)\right\|_{\mathrm{F}}\mathbf{I}_2\succeq\left\|\nabla^2\Gamma_{k,i}(\mathbf{u}_m)\right\|_2\mathbf{I}_2\succeq\nabla^2\Gamma_{k,i}(\mathbf{u}_m)$, the value of $\delta_{k,i}$ can be determined as $\delta_{k,i}=\left\|\nabla^2\Gamma_{k,i}(\mathbf{u}_m)\right\|_{\mathrm{F}}=\left[\left(\frac{\partial^2\Gamma_{k,i}(\mathbf{u}_m)}{\partial x_m^2}\right)^2+\right.\\ \left.\left(\frac{\partial^2\Gamma_{k,i}(\mathbf{u}_m)}{\partial x_m\partial y_m}\right)^2+\left(\frac{\partial^2\Gamma_{k,i}(\mathbf{u}_m)}{\partial y_m\partial x_m}\right)^2+\left(\frac{\partial^2\Gamma_{k,i}(\mathbf{u}_m)}{\partial y_m^2}\right)^2\right]^{\frac{1}{2}}$.
For addressing non-convex constraint (7d), let us define $\Upsilon_{k,i}(\mathbf{u}_m)=\sum_{s(j)>s(k)}\left|\mathbf{h}_i^{\mathrm{H}}\left(\tilde{\mathbf{u}}\right)\mathbf{w}_j\right|^2+\sigma_i^2$. Similarly, based on Taylor's theorem, we can construct a convex upper bound on $\Upsilon_{k,i}(\mathbf{u}_m)$, i.e., the RHS of (7d), denoted as
\begin{align}
	\Upsilon_{k,i}\left(\mathbf{u}_m\right)\leq&\Upsilon_{k,i}\Big(\mathbf{u}_m^t\Big)+\nabla\Upsilon_{k,i}\Big(\mathbf{u}_m^t\Big)^\mathrm{T}\Big(\mathbf{u}_m-\mathbf{u}_m^t\Big)\notag\\
	&+\frac{\psi_{k,i}}{2}\Big(\mathbf{u}_m-\mathbf{u}_m^t\Big)^{\mathrm{T}}\Big(\mathbf{u}_m-\mathbf{u}_m^t\Big)\notag\\
	\triangleq&\Upsilon_{k,i}^{\mathrm{ub},t}\left(\mathbf{u}_m\right),
\end{align}
where the gradient and Hessian matrix of $\Upsilon_{k,i}(\mathbf{u}_m)$ over $\mathbf{u}_m$ can be respectively given by
\begin{align}
	&\nabla\Upsilon_{k,i}\left(\mathbf{u}_m\right)=\sum_{s(j)>s(k)}\nabla\Gamma_{j,i}\left(\mathbf{u}_m\right),\\
	&\nabla^2\Upsilon_{k,i}\left(\mathbf{u}_m\right)=\sum_{s(j)>s(k)}\nabla^2\Gamma_{j,i}\left(\mathbf{u}_m\right),
\end{align}
and $\psi_{k,i}$ can be determined as $\psi_{k,i}=\sum_{s(j)>s(k)}\delta_{j,i}$ to satisfy $\psi_{k,i}\mathbf{I}_2\succeq\nabla^2\Upsilon_{k,i}(\mathbf{u}_m)$. 

Hereto, the remaining difficulty to tackling problem (12) lies in handling non-convex constraint (6c). Note that constraint (6c) is equivalent to $\|\mathbf{u}_m-\mathbf{u}_n\|_2^2\geq D^2, \forall m\neq n$. Since the term $\|\mathbf{u}_m-\mathbf{u}_n\|_2^2$ is convex with respect to $\mathbf{u}_m$, by applying the first-order Taylor series expansion, it is lower bounded by
\begin{align}
	\big\|\mathbf{u}_m-\mathbf{u}_n\big\|_2^2\geq&\big\|\mathbf{u}_m^t-\mathbf{u}_n\big\|_2^2+2\Big(\mathbf{u}_m^t-\mathbf{u}_n\Big)^{\mathrm{T}}\Big(\mathbf{u}_m-\mathbf{u}_m^t\Big)\notag\\
	\triangleq&U^{\mathrm{lb},t}_n\left(\mathbf{u}_m\right).
\end{align}
With (13), (14), and (17), in the $t$-th iteration of SCA, the optimization problem of the $m$-th MA's position $\mathbf{u}_m$ can be transformed into
\begin{subequations}
	\begin{align}
		\max_{\mathbf{u}_m,\{R_{k\rightarrow i},\alpha_{k,i},\beta_{k,i}\}}& \quad \sum_{k=1}^{K}R_{k\rightarrow k}\\
		\mathrm{s.t.} & \quad \frac{1}{\alpha_{k,i}}\leq\Gamma_{k,i}^{\mathrm{lb},t}\left(\mathbf{u}_m\right),~\forall k,i\in\mathcal{K},\\
		& \quad \beta_{k,i}\geq \Upsilon_{k,i}^{\mathrm{ub},t}\left(\mathbf{u}_m\right),~\forall k,i\in\mathcal{K},\\
		& \quad U^{\mathrm{lb},t}_n\left(\mathbf{u}_m\right)\geq D^2,~\forall m\neq n,\\
		& \quad \mathrm{(6b),(6d),(7e),(10)},\notag
	\end{align}
\end{subequations}
which is a convex problem and thus can be efficiently solved via the CVX toolbox\cite{ref32}. 

\begin{figure}[t]
	\vspace{-0.3cm}
	\begin{algorithm}[H]
		\caption{Alternating Optimization Algorithm for Solving Problem (7)}
		\begin{algorithmic}[1]
			\STATE Choose a decoding order $\left\{s(k)\right\}$; initialize feasible solutions $\left\{\{\mathbf{w}_k^0\},\{\mathbf{u}_m^0\}\right\}$ to problem (7); set $t=0$.
			
			\REPEAT
			
			\STATE Obtain $\left\{\mathbf{w}_k^{t+1}\right\}$ by solving problem (11). 
			
			\FOR{$m = 1: M$}
			\STATE Compute $\left\{\nabla\Gamma_{k,i}(\mathbf{u}_m^t),\nabla\Upsilon_{k,i}(\mathbf{u}_m^t),\delta_{k,i},\psi_{k,i}\right\}$.
			\STATE Obtain $\mathbf{u}_m^{t+1}$ by solving problem (18). 
			\ENDFOR
			\STATE $t\leftarrow t+1$.
			
			\UNTIL{the fractional increase of the objective value is less than a threshold $\epsilon>0$.}
			
		\end{algorithmic}
		\label{alg1} 
	\end{algorithm}
	\vspace{-0.5cm}
\end{figure}
\textit{2) Overall algorithm:} Based on the above subproblems, the overall algorithm
for solving problem (7) is summarized in Algorithm 1.
In lines 3-7, the beamforming vectors and each MA position are alternately optimized. Since the objective value of problem (7) is non-decreasing over iterations and the solution set is compact, the convergence of Algorithm 1 is guaranteed. Besides, the computational complexity of this algorithm is analyzed as follows. In line 3, the computational complexity for solving problem (11) is $\mathcal{O}\left(\max\left(M,3K^2+2K\right)^4M^{1/2}\ln\frac{1}{\varepsilon}\right)$, where $\varepsilon>0$ denotes the solution accuracy\cite{ref38}. The computational complexities of lines 5-6 are $\mathcal{O}\left(K^2L_k^2\right)$ and $\mathcal{O}\left(\left(3K^2+M\right)^{1.5}\ln\frac{1}{\varepsilon}\right)$, respectively. Therefore, the overall computational complexity of Algorithm 1 is $\mathcal{O}\left(T_{\max}\left(\max\left(M,3K^2+2K\right)^4M^{1/2}\ln\frac{1}{\varepsilon}\right.\right.$
$\left.\left.+M\left(K^2L_k^2+M\left(3K^2+M\right)^{1.5}\ln\frac{1}{\varepsilon}\right)\right)\right)$, where $T_{\max}$ denotes the maximum number of iterations.

\section{Simulation Results}
In this section, simulation results are provided to evaluate the performance of our proposed MA array-enhanced downlink NOMA system. We employ the geometry channel model\cite{ref3}, in which the numbers of  transmit channel paths from the BS to all users are identical, i.e., $L_k=L,\forall k\in\mathcal{K}$. The PRV for each user is modeled as a circularly symmetric complex Gaussian (CSCG) random vector with its element satisfying $f_{k,\ell}\sim\mathcal{CN}\left(0,\rho d_k^{-\alpha}/L\right),1\leq \ell\leq L$, where $\rho=-30\text{~dB}$ denotes the path loss at the reference distance, i.e., 1 meter (m), $d_k$ denotes
the distance from the BS to user $k$, and $\alpha=2.8$ represents the path loss exponent. The elevation
and azimuth AoDs of the transmit channel paths from the BS to each user are presumed to be independent and identically variables, which follow uniform distributions over $[0,\pi]$. The distance between the BS and each user are randomly generated from 50 to 100 m. Besides, the side length of the MA moving region is set as $A=3\lambda$, the minimum inter-MA distance is set to $D=\lambda/2$, the average power of the AWGN is set as $\sigma_k^2=-80\text{~dBm},\forall k\in\mathcal{K}$, and the minimum rate requirement for each user is set as $R_{k}^{\min}=0.25\text{~bps/Hz}$. All points in our simulation curves is averaged over 1000 independent channel realizations.

\begin{figure}[!t]
	\vspace{-0.2cm}
	\centering
	\includegraphics[width=0.43\textwidth]{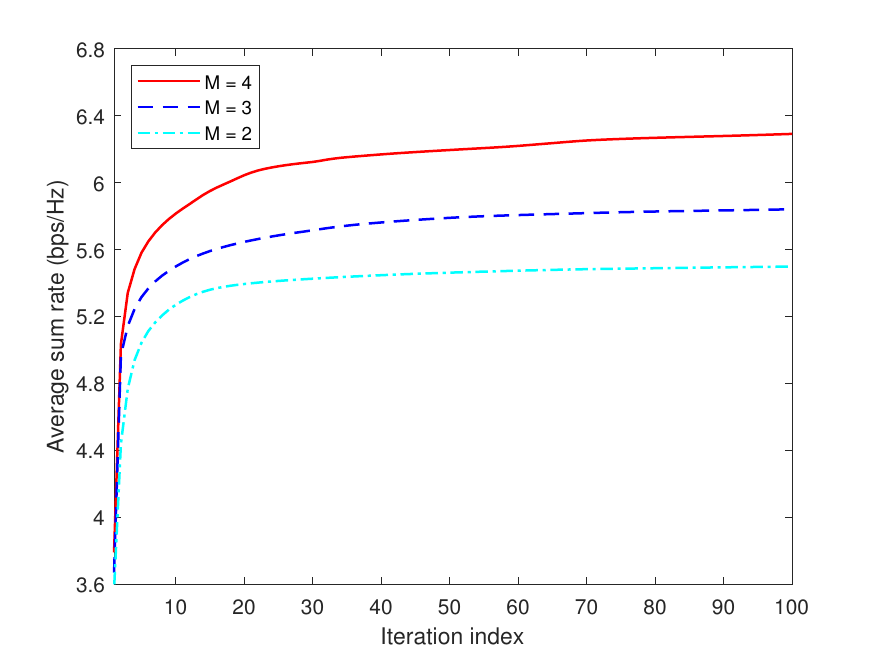}
	\caption{Evaluation of the convergence of our proposed algorithm for $K=3$ and $P_s=10\text{~dBm}$.}
	\label{convergence_performance}
\end{figure}
In Fig. 2, we illustrate the convergence performance of our proposed Algorithm 1, where the number of antennas at the BS is set as $M=2,3,4$, the number of users is set as $K=3$, and the maximum transmit power of the BS is set as $P_s=10\text{ dBm}$. It can be observed that the sum rate obtained by our proposed algorithm increases monotonically with the iteration index and converges to a constant after 90 iterations on average. 


\begin{figure}[!t]
	\centering
	\includegraphics[width=0.43\textwidth]{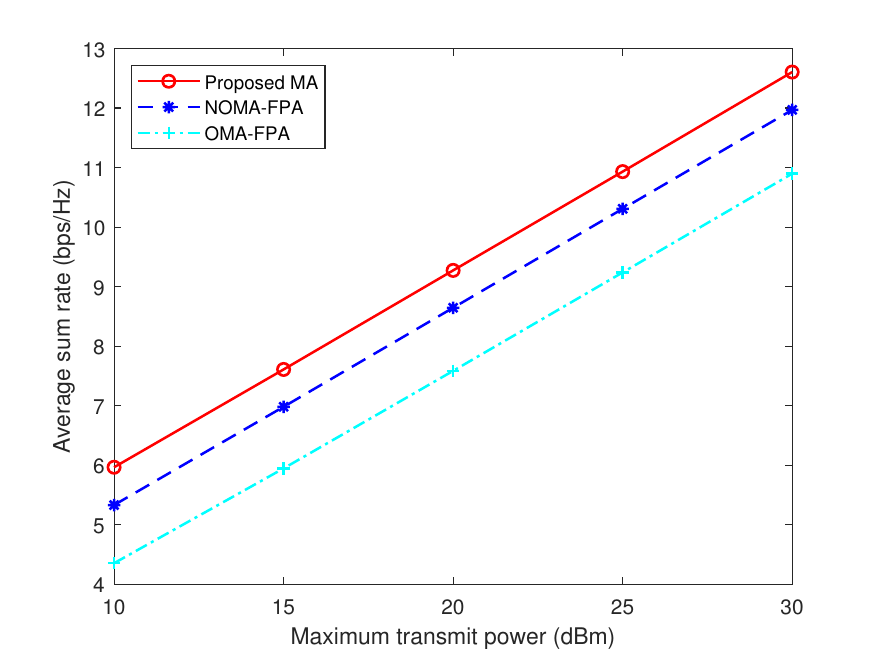}
	\caption{Average sum rate of different schemes versus the maximum transmit power $P_s$ for $M=4$ and $K=3$.}
	\label{sum_rate_vs_Ps}
\end{figure}
Next, to demonstrate the superiority of deploying the MAs, we compare the users' sum rate of the proposed algorithm with other benchmark schemes. For the ``NOMA-FPA'' scheme, we consider that the BS is equipped with an FPA-based uniform planar array with $M$ antennas, spaced by $\lambda/2$, and serves all users by NOMA. For the ``OMA-FPA'' scheme, the BS is equipped with an FPA-based uniform planar array and serves all users via equal-length time division multiple access slots.

Fig. 3 illustrates the sum rate of different schemes versus the maximum transmit power, where the number of antennas at the BS is set as $M=4$ and the number of users is set as $K=3$. First, it is observed that the sum rate of all considered schemes increases with the maximum transmit power. This is because the SINR of each user can be improved by increasing the transmit power. Besides, to achieve the same sum rate, our proposed scheme requires less transmit power than other benchmark schemes. For instance, to obtain a sum rate of 9 bps/Hz, our proposed scheme requires only about 19 dBm transmit power, while the NOMA-FPA scheme and the OMA-FPA scheme need about 21 dBm and 24 dBm, respectively. The reason is that integrating MA into NOMA systems can not only improve channel conditions but also magnify channel disparities among users, thereby significantly enhancing the rate advantage of NOMA over conventional OMA. 


\begin{figure}[!t]
	\vspace{-0.2cm}
	\centering
	\includegraphics[width=0.43\textwidth]{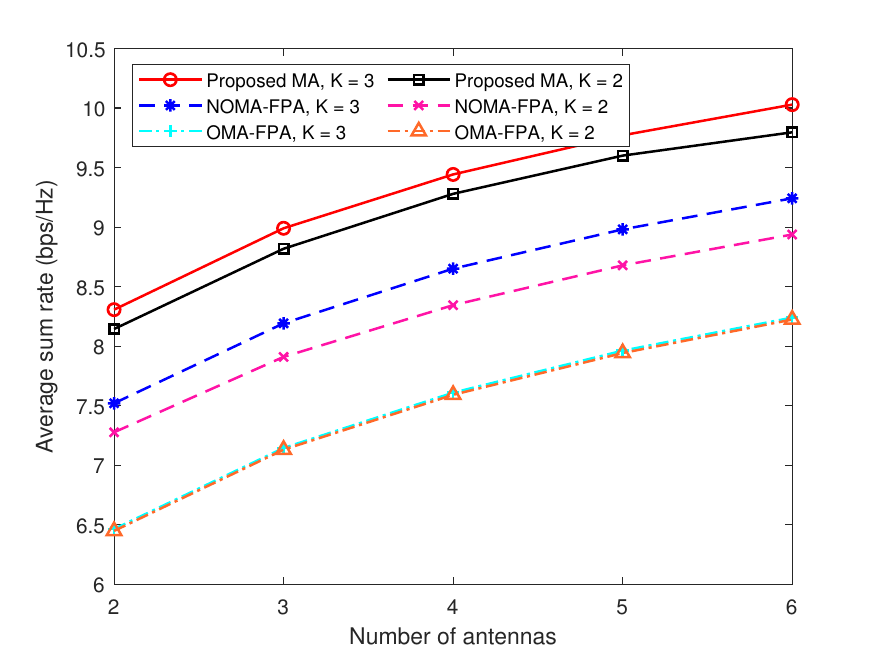}
	\caption{Average sum rate of different schemes versus the number of antennas $M$ for $K=2,3$ and $P_s=20\text{~dBm}$.}
	\label{sum_rate_vs_M}
\end{figure}
In Fig. 4, we show the sum rate of different schemes versus the number of antennas at the BS, where the number of users is set as $K=2,3$ and the maximum transmit power is set as $P_s=20\text{~dBm}$. As can be observed, the sum rate of all schemes increases with the number of antennas due to the enhanced spatial diversity and beamforming gains. However, with increasing the user number, the sum rate of the OMA-FPA scheme remains unchanged since OMA only serves users via orthogonal channel resources. On the contrary, the sum rates of the NOMA-based schemes increase with the user number thanks to the multiplexing gain in power domain. Besides, it can be observed that our proposed scheme outperforms other benchmark schemes under any value of $M$ and $K$. 


\section{Conclusion}
This paper investigated the MA-enhanced downlink NOMA system, where a BS equipped with an MA array serves multiple single-FPA users. Specifically, an optimization problem was formulated for maximizing the users' sum rate by jointly optimizing the positions of MAs and the beamforming vectors of the BS. To tackle this sophisticated non-convex problem, an efficient algorithm was developed by leveraging alternating optimization and SCA techniques. Simulation results showed that compared to FPA systems and traditional OMA systems, our proposed solution for MA-enhanced NOMA systems can provide remarkable performance improvements by effectively exploiting spatial DoFs via antenna position optimization.

{\appendices
	\section{Derivations of $\nabla\Gamma_{k,i}(\mathbf{u}_m)$ and $\nabla^2\Gamma_{k,i}(\mathbf{u}_m)$}
		\begin{figure*}[!t]
		\small
		\setcounter{equation}{20}
		\begin{subequations}
			\begin{align}
				\frac{\partial^2\Gamma_{k,i}(\mathbf{u}_m)}{\partial x_m^2}=&-\frac{4\pi^2}{\lambda^2}\sum_{\ell_1=1}^{L_k}\sum_{\ell_2=1}^{L_k}\left|\left[\mathbf{B}_{k,i,m}\right]_{\ell_1,\ell_2}\right|\left(-\sin\theta_{i,\ell_1}\cos\phi_{i,\ell_1}+\sin\theta_{i,\ell_2}\cos\phi_{i,\ell_2}\right)^2\cos(\bar{\omega}_{k,i,m,\ell_1,\ell_2}(\mathbf{u}_m))\notag\\
				&-\frac{4\pi^2}{\lambda^2}\sum_{\ell_3=1}^{L_k}\left|\left[\mathbf{c}_{k,i,m}\right]_{\ell_3}\right|\sin^2\theta_{i,\ell_3}\cos^2\phi_{i,\ell_3}\cos(\bar{\kappa}_{k,i,m,\ell_3}(\mathbf{u}_m)),\\
				\frac{\partial^2\Gamma_{k,i}(\mathbf{u}_m)}{\partial x_my_m}=&\frac{\partial^2\Gamma_{k,i}(\mathbf{u}_m)}{\partial y_mx_m}=-\frac{4\pi^2}{\lambda^2}\sum_{\ell_1=1}^{L_k}\sum_{\ell_2=1}^{L_k}\left|\left[\mathbf{B}_{k,i,m}\right]_{\ell_1,\ell_2}\right|\left(-\sin\theta_{i,\ell_1}\cos\phi_{i,\ell_1}+\sin\theta_{i,\ell_2}\cos\phi_{i,\ell_2}\right)\left(-\cos\theta_{i,\ell_1}+\cos\theta_{i,\ell_2}\right)\notag\\
				&\times\cos(\bar{\omega}_{k,i,m,\ell_1,\ell_2}(\mathbf{u}_m))
				-\frac{4\pi^2}{\lambda^2}\sum_{\ell_3=1}^{L_k}\left|\left[\mathbf{c}_{k,i,m}\right]_{\ell_3}\right|\sin\theta_{i,\ell_3}\cos\phi_{i,\ell_3}\cos\theta_{i,\ell_3}\cos(\bar{\kappa}_{k,i,m,\ell_3}(\mathbf{u}_m)),\\
				\frac{\partial^2\Gamma_{k,i}(\mathbf{u}_m)}{\partial y_m^2}=&-\frac{4\pi^2}{\lambda^2}\sum_{\ell_1=1}^{L_k}\sum_{\ell_2=1}^{L_k}\left|\left[\mathbf{B}_{k,i,m}\right]_{\ell_1,\ell_2}\right|\left(-\cos\theta_{i,\ell_1}+\cos\theta_{i,\ell_2}\right)^2\cos(\bar{\omega}_{k,i,m,\ell_1,\ell_2}(\mathbf{u}_m))\notag\\
				&-\frac{4\pi^2}{\lambda^2}\sum_{\ell_3=1}^{L_k}\left|\left[\mathbf{c}_{k,i,m}\right]_{\ell_3}\right|\cos^2\theta_{i,\ell_3}\sin(\bar{\kappa}_{k,i,m,\ell_3}(\mathbf{u}_m)).
			\end{align}
		\end{subequations}
		\hrulefill
		\vspace{-0.2cm}
	\end{figure*}
	Recall that $\mathbf{h}_k(\tilde{\mathbf{u}})=\mathbf{G}_k^{\mathrm{H}}(\tilde{\mathbf{u}})\mathbf{f}_k$. Let $w_{k,n}$ denote the $n$-th element of $\mathbf{w}_k$ and define $\zeta_{k,i,m}=\sum_{n\neq m}\mathbf{f}_i^{\mathrm{H}}\mathbf{g}_i(\mathbf{u}_n)w_{k,n}$. Armed with these, $\left|\mathbf{h}_i^{\mathrm{H}}\left(\tilde{\mathbf{u}}\right)\mathbf{w}_k\right|^2$ can be expanded as
	\begin{align}
		\setcounter{equation}{18}
		&\Gamma_{k,i}\left(\mathbf{u}_m\right)=\left|\mathbf{h}_i^{\mathrm{H}}\left(\tilde{\mathbf{u}}\right)\mathbf{w}_k\right|^2=\left|\mathbf{f}_i^{\mathrm{H}}\mathbf{g}_i(\mathbf{u}_m)w_{k,m}+\zeta_{k,i,m}\right|^2\notag\\
		&=\sum_{\ell_1=1}^{L_i}\sum_{\ell_2=1}^{L_i}\left|\left[\mathbf{B}_{k,i,m}\right]_{\ell_1,\ell_2}\right|\cos\big(\bar{\omega}_{k,i,m,\ell_1,\ell_2}(\mathbf{u}_m)\big)\notag\\
		&+\sum_{\ell_3=1}^{L_i}\left|\left[\mathbf{c}_{k,i,m}\right]_{\ell_3}\right|\cos\big(\bar{\kappa}_{k,i,m,\ell_3}(\mathbf{u}_m)\big)+\left|\zeta_{k,i,m}\right|^2,
	\end{align}
	where $\mathbf{B}_{k,i,m}\triangleq\left|w_{k,m}\right|^2\mathbf{f}_i\mathbf{f}_i^{\mathrm{H}}$,  $\mathbf{c}_{k,i,m}\triangleq2w_{k,m}^*\zeta_{k,i,m}\mathbf{f}_i$, $\bar{\omega}_{k,i,m,\ell_1,\ell_2}\left(\mathbf{u}_m\right)\triangleq\mathrm{arg}\left(\left[\mathbf{B}_{k,i,m}\right]_{\ell_1,\ell_2}\right)+\frac{2\pi}{\lambda}\big(-\rho_{i,\ell_1}\left(\mathbf{u}_m\right)+\\\rho_{i,\ell_2}\left(\mathbf{u}_m\right)\big)$, and $\bar{\kappa}_{k,i,m,\ell_3}(\mathbf{u}_m)\triangleq-\mathrm{arg}\left(\left[\mathbf{c}_{k,i,m}\right]_{\ell_1,\ell_2}\right)+\frac{2\pi}{\lambda}\\\rho_{i,\ell_3}(\mathbf{u}_m)$. Therefore, the gradient vector of $\Gamma_{k,i}(\mathbf{u}_m)$ with respect to $\mathbf{u}_m$, i.e.,  $\nabla\Gamma_{k,i}(\mathbf{u}_m)=\left[\frac{\partial\Gamma_{k,i}(\mathbf{u}_m)}{\partial x_m},\frac{\partial\Gamma_{k,i}(\mathbf{u}_m)}{\partial y_m}\right]^{\mathrm{T}}$, can be derived in a closed form as
	\begin{subequations}
		\small
		\begin{align}
			&\frac{\partial\Gamma_{k,i}(\mathbf{u}_m)}{\partial x_m}=-\frac{2\pi}{\lambda}\sum_{\ell_1=1}^{L_k}\sum_{\ell_2=1}^{L_k}\left|\left[\mathbf{B}_{k,i,m}\right]_{\ell_1,\ell_2}\right|\sin(\bar{\omega}_{k,i,m,\ell_1,\ell_2}(\mathbf{u}_m))\notag\\
			&\times\left(-\sin\theta_{i,\ell_1}\cos\phi_{i,\ell_1}+\sin\theta_{i,\ell_2}\cos\phi_{i,\ell_2}\right)\notag\\
			&-\frac{2\pi}{\lambda}\sum_{\ell_3=1}^{L_k}\left|\left[\mathbf{c}_{k,i,m}\right]_{\ell_3}\right|\sin\theta_{i,\ell_3}\cos\phi_{i,\ell_3}\sin(\bar{\kappa}_{k,i,m,\ell_3}(\mathbf{u}_m)),\\
			&\frac{\partial\Gamma_{k,i}(\mathbf{u}_m)}{\partial y_m}=-\frac{2\pi}{\lambda}\sum_{\ell_1=1}^{L_k}\sum_{\ell_2=1}^{L_k}\left|\left[\mathbf{B}_{k,i,m}\right]_{\ell_1,\ell_2}\right|\sin(\bar{\omega}_{k,i,m,\ell_1,\ell_2}(\mathbf{u}_m))\notag\\
			&\times\left(-\cos\theta_{i,\ell_1}+\cos\theta_{i,\ell_2}\right)\notag\\
			&-\frac{2\pi}{\lambda}\sum_{\ell_3=1}^{L_k}\left|\left[\mathbf{c}_{k,i,m}\right]_{\ell_3}\right|\cos\theta_{i,\ell_3}\sin(\bar{\kappa}_{k,i,m,\ell_3}(\mathbf{u}_m)).
		\end{align}
	\end{subequations}
	Similarly, the Hessian matrix of $\Gamma_{k,i}(\mathbf{u}_m)$ with respect to $\mathbf{u}_m$, i.e., $\nabla^2\Gamma_{k,i}(\mathbf{u}_m)=\begin{bmatrix}
	\frac{\partial^2\Gamma_{k,i}(\mathbf{u}_m)}{\partial x_m^2},\frac{\partial^2\Gamma_{k,i}(\mathbf{u}_m)}{\partial x_m\partial y_m}\\
	\frac{\partial^2\Gamma_{k,i}(\mathbf{u}_m)}{\partial y_m\partial x_m},\frac{\partial^2\Gamma_{k,i}(\mathbf{u}_m)}{\partial y_m^2}
	\end{bmatrix}$, can also be derived in a closed form, with the expression given in (21), shown at the top of this page.
}

\bibliography{reference}

\begin{thebibliography}{10}
\providecommand{\url}[1]{#1}
\csname url@samestyle\endcsname
\providecommand{\newblock}{\relax}
\providecommand{\bibinfo}[2]{#2}
\providecommand{\BIBentrySTDinterwordspacing}{\spaceskip=0pt\relax}
\providecommand{\BIBentryALTinterwordstretchfactor}{4}
\providecommand{\BIBentryALTinterwordspacing}{\spaceskip=\fontdimen2\font plus
\BIBentryALTinterwordstretchfactor\fontdimen3\font minus
  \fontdimen4\font\relax}
\providecommand{\BIBforeignlanguage}[2]{{%
\expandafter\ifx\csname l@#1\endcsname\relax
\typeout{** WARNING: IEEEtran.bst: No hyphenation pattern has been}%
\typeout{** loaded for the language `#1'. Using the pattern for}%
\typeout{** the default language instead.}%
\else
\language=\csname l@#1\endcsname
\fi
#2}}
\providecommand{\BIBdecl}{\relax}
\BIBdecl

\bibitem{ref1}
L.~Zhu, W.~Ma, and R.~Zhang, ``Movable antennas for wireless communication:
  Opportunities and challenges,'' \emph{IEEE Commun. Mag.}, vol.~62, no.~6, pp.
  114--120, Jun. 2024.

\bibitem{ref3}
------, ``Modeling and performance analysis for movable antenna enabled
  wireless communications,'' \emph{IEEE Trans. Wireless Commun.}, vol.~23,
  no.~6, pp. 6234--6250, Jun. 2024.

\bibitem{ref35}
B.~Ning \emph{et~al.}, ``Movable antenna-enhanced wireless communications:
  General architectures and implementation methods,''
  \textit{arXiv:2407.15448}, 2024.

\bibitem{ref27}
W.~Mei \emph{et~al.}, ``Movable-antenna position optimization: A graph-based
  approach,'' \emph{IEEE Wireless Commun. Lett.}, vol.~13, no.~7, pp.
  1853--1857, Jul. 2024.

\bibitem{ref9}
L.~Zhu \emph{et~al.}, ``Movable-antenna enhanced multiuser communication via
  antenna position optimization,'' \emph{IEEE Trans. Wireless Commun.},
  vol.~23, no.~7, pp. 7214--7229, Jul. 2024.

\bibitem{ref39}
Y.~Gao, Q.~Wu, and W.~Chen, ``Joint transmitter and receiver design for movable
  antenna enhanced multicast communications,'' \emph{IEEE Trans. Wireless
  Commun.}, vol.~23, no.~12, pp. 18\,186--18\,200, 2024.

\bibitem{ref36}
N.~Li \emph{et~al.}, ``Over-the-air computation via {2-D} movable antenna
  array,'' \emph{IEEE Wireless Commun. Lett.}, vol.~14, no.~1, pp. 33--37,
  2025.

\bibitem{ref43}
H.~Mao \emph{et~al.}, ``Sum rate maximization for movable antenna enhanced
  multiuser covert communications,'' \emph{IEEE Wireless Commun. Lett.}, 2024,
  early access.

\bibitem{ref37}
J.~Chen \emph{et~al.}, ``Low-complexity beamforming design for {RIS}-assisted
  fluid antenna systems,'' in \emph{Proc. IEEE ICC Workshops}, 2024, pp.
  1377--1382.

\bibitem{ref38}
X.~Mu \emph{et~al.}, ``Exploiting intelligent reflecting surfaces in {NOMA}
  networks: Joint beamforming optimization,'' \emph{IEEE Trans. Wireless
  Commun.}, vol.~19, no.~10, pp. 6884--6898, 2020.

\bibitem{ref41}
Y.~Zhou \emph{et~al.}, ``Movable antenna empowered downlink {NOMA} systems:
  Power allocation and antenna position optimization,'' \emph{IEEE Wireless
  Commun. Lett.}, vol.~13, no.~10, pp. 2772--2776, 2024.

\bibitem{ref2}
N.~Li, P.~Wu, B.~Ning, and L.~Zhu, ``Sum rate maximization for movable antenna
  enabled uplink {NOMA},'' \emph{IEEE Wireless Commun. Lett.}, vol.~13, no.~8,
  pp. 2140--2144, Aug. 2024.

\bibitem{ref42}
Z.~Xiao \emph{et~al.}, ``Movable antenna aided {NOMA}: Joint antenna
  positioning, precoding, and decoding design,'' \textit{arXiv:2412.12531},
  2024.

\bibitem{ref32}
S.~Boyd and L.~Vandenberghe, \emph{Convex optimization}.\hskip 1em plus 0.5em
  minus 0.4em\relax Cambridge unive-\\rsity press, 2004.

\end{thebibliography}
\bibliographystyle{IEEEtran}

\end{document}